# A NEW LOCALIZED NETWORK BASED ROUTING MODEL IN COMPUTER AND COMMUNICATION NETWORKS

Abdulbaset H. Mohammad

School of Computing, Informatics and Media, University of Bradford, Bradford, United Kingdom

a.h.t.mohammad@bradford.ac.uk

***ABSTRACT***

*In view of the fact that routing algorithms are network layer entities and the varying performance of any routing algorithm depends on the underlying networks. Localized routing algorithms avoid the problems associated with the maintenance of global network state by using statistics of flow blocking probabilities. We developed a new network parameter that can be used to predict which network topology gives better performance on the quality of localized QoS routing algorithms. Using this parameter we explore a simple model that can be rewired to introduce increasing the performance. We find that this model have small characteristic path length. Simulations of random and complex networks used to show that the performance is significantly affected by the level of connectivity.*

## 1. INTRODUCTION

The constantly changing nature of the Internet presents a challenge in evaluating routing behaviour in large-scale networks. Various studies have evaluated the performance of different routing algorithms on simulated networks with various topologies, such as ISP [1], random networks [3]. However, due to the characteristics and decentralized nature of the Internet it makes it hard to define a typical topology [5] [6]. In view of the fact that routing algorithms are network layer entities and the varying performance of any routing algorithm depends on the underlying network, in the efficient design and assessment of the performance of routing algorithms it is necessary to use different structures which vary in their characteristics and parameters. The study of complex networks has shown that they are heavily depending on the network's structure [7].

However, the global QoS routing algorithms needs providing up-to-date changes of all links at all times making them impractical [2]. Such high levels of exchange may incur large communication, processing overheads and affect QoS routing algorithms [4]. An alternative of global QoS routing algorithms is eliminating typical link state advertisements [8] [12].

In this paper, we focus on localized QoS routing algorithms. The localized QoS routing proposed [8] [9] [10] is attempts to overcome the problems associated with the maintenance of the global network state information by making routing decisions based solely on the information collected locally at each source. In localized QoS routing schemes each source node has a predetermined set of candidate paths to each of the destinations. It has been shown that localized routing algorithms discriminate against alternative paths which prefer the shortest paths. These minimum hop algorithms usually outperform algorithms that do not take path length into consideration [11] [13]. Biasing towards short paths is particularly attractive in large-scale networks, since path length is a relatively stable metric compared with link delay [14].

In this paper we developed a new network parameter that can be used to predict which network topology gives better performance on the quality of localized QoS routing algorithms. Using this parameter we explore a simple model that can be rewired to introduce increasing the connections that the network accepts. We find that this model have small characteristic path length. Simulations of random and complex networks used to show that the performance is significantly affected by the level of connectivity.

## 2. Related Work

The ubiquitous nature of complex networks has been recently studied in many fields in science [15] [16 [17], such as the Internet, the power grid and the World Wide Web (WWW). These have naturally been concerned with the network structure and connectivity constraints.

### 2.1. Complex Networks

Three well-known properties in complex network topologies are *random networks*, the *small-world network* and *scale-free network* properties. These networks are intensively studied and fairly well developed, each of these networks are characterized by the way in which networks are generated and by several statistical metrics.

#### 2.1.1. Random Networks

The most investigated model of random networks is the binomial random networks. The general discovery of this model was that many properties of these networks like appearance of trees and cycles arise quite suddenly at a threshold value [19].

However, despite the fact that the position of the edges is random in random networks, a typical random network is rather homogeneous, the majority of the nodes having the same number of edges.

#### 2.1.2. Small World Networks

The small world networks stands for networks with a small average path length and a high clustering coefficients, while the scale free networks stands for a network in which a few nodes have a very large degree whereas most nodes have a small degree. The *small world* networks are characterized by two main features [18]. Firstly, its average path length $\bar{h}$ is at most logarithmic in the number of nodes. Secondly, it has a high *clustering coefficient C*, which is the likeliness that if a node *a* is linked to *b* and *b* is linked to *c* then *a* is also linked to *c*. Thus, $\bar{h}$ and *C are* indicators of the network. A *scale-free* network is characterized by a distribution of degree that follows a power-law. If $p(x)$ denotes the fraction of nodes having degree *x*, then the network is *scale-free* if $p(x) = cx^{-\alpha}$ , where *c* is a constant [19].

Although, various topologies manifest obvious small world properties, it has been shown that small world networks and scale-free topologies are rewiring randomly and they are in fact resulting in very sparsely connected networks. However, their topologies are not balancing the connectivity between nodes and it is inadequate for the performance of localized routing algorithms.

### 2.2. Localized QoS Routing Algorithms

Localized Quality of Service routing has recently been introduced as a new approach in the context of QoS routing. To the best of our knowledge our simulation study is the first that considers network based model on the performance of localized QoS routing. The main localized quality of service routing algorithms are:

### 2.2.1. Localized Proportional Sticky Routing Algorithm

The localized proportional sticky routing algorithm (PSR) [8] was the first localized QoS routing scheme used in the context of computer networks. The basic idea behind the PSR approach assumes that route level statistics, such as the number of flows blocked, is the only available QoS state information at a source and based on this information the algorithm attempts to proportionally distribute the traffic load from a source to a destination among the set of candidate paths, according to their flow blocking probability. With this scheme each source node needs to maintain a set of candidate paths $R$. A path is based on flow blocking probability and the load is proportionally distributed to the destination among the predefined paths. In PSR there are minimum hop paths $R^{min}$ and alternative paths $R^{alt}$, where $R = R^{min} \cup R^{alt}$.

The PSR algorithm can be viewed as operating in two stages: proportional flow routing and computation of flow proportions. The scheme proceeds in cycles of variable lengths which form an observation period. During each cycle along a path $r$, any incoming connection request can be routed among paths selected from a set of eligible paths $R^{alt}$, which initially may include all candidate paths. A candidate path is ineligible depending on the maximum permissible flow blocking parameter $\gamma_r$, which determines how many times this candidate path can block a connection request before it becomes ineligible.

For each minimum hop path, $\gamma_r$ is set to $\hat{y}$, which is a configurable parameter, whereas the alternative path $\gamma_r$ is dynamically adjusted between 1 and $\hat{y}$. When all candidate paths become ineligible a cycle ends and all parameters are reset to start the next cycle. An eligible path is finally selected depending on its flow proportions. The larger the flow proportions, the larger chances for selection.

At the end of the observation period, a new flow proportion $\alpha_r$ is computed for each path in the candidate path set, based on its observed blocking probability $b_r$. After each observation period the minimum hop path flow proportions are adjusted to equalize their blocking probability ($\alpha_r.b_r$). For the alternative paths, the minimum blocking probability among the minimum hop paths b* is used to control their flow proportion. That is, for each $r \in R^{alt}$, if $b_r < \psi b^*$, $\gamma_r$=min ($\gamma_r$+1, $\hat{y}$). If $b_r > b^*$, $\gamma_r$=max ($\gamma_r$-1, 1), where $\psi$ is a configurable parameter to limit the 'knock-on' effect [3] under system overloads. Note that $\gamma_r \geq 1$ ensures that some flows are routed along alternative paths to measure their quality.

### 2.2.2. Localized Credit Based Routing Algorithm

The Credit Based Routing (CBR) [9] algorithm uses a simple routing procedure to route traffic across the network. The CBR scheme performs routing using crediting scheme for each candidate path that rewards a path upon flow acceptance and penalizes it upon flow rejection. The larger path credits, the larger chances for selection. The CBR algorithm keeps updating each path's credit upon flow acceptance and rejection and it does not compute a flow proportion. It is also keeps monitoring the flow blocking probabilities for each path and conveys the data to the credit scheme to use it in path to path selection. A set of candidate paths R between each source and destination is required in the CBR algorithm. Like PSR, CBR predetermined a minimum hop set $R^{\min}$ and an alternative paths set $R^{alt}$ where $R = R^{\min} \cup R^{alt}$ . CBR selects the largest credit path P.credits in each set, minimum hop paths set $R^{\min}$ and alternative paths set $R^{alt}$ upon flow arrival. The flow is routed along the minimum hop path that has the largest credit $P^{\min}$ which is larger than the alternative path that has the largest credits $P^{alt}$; the flow is routed along an alternative path using this formula (1):

$$P^{\min}.credits \geq \Phi \times P^{alt}.credits \text{ , where } \Phi \leq 1 \qquad (2)$$

$\Phi$ is a system parameter that controls the usage of alternative paths. The CBR uses blocking probability in crediting schemes to improve the performance of the algorithm. The path credits are incremented or decremented upon flow acceptance or rejection using statistics of the path blocking probability.

However, CBR uses a MAX_CREDITS system parameter to determine the maximum attainable credits for each path by computing the blocking probability.

$$0 \leq Credits \leq MAX_CREDITS \qquad (3)$$

CBR algorithm records rejection and acceptance for each path and uses a moving window for a predetermined period of M connection requests. It uses 1 for flow acceptance and 0 for flow rejection, dividing the number of 0's by M to calculate each path blocking probability for the period of M connection requests.

### 2.2.3. Localized Quality Based Routing Algorithm

We also consider the Quality Based Routing (QBR) proposed [20], QBR monitors the current residual bandwidth for each path on its bottleneck links, and incorporates its values into simple average path qualities.

When a flow is accepted and the value of residual bandwidth is greater than the value of requested bandwidth, P.Quality is set to one. When a flow is accepted, but the residual bandwidth of the path is less than requested bandwidth, P.Quality is set to a value less than one.

On the other hand, if the flow is rejected, P.Quality is set to a value less than zero. Hence, QBR continuously updates the resulting normalized values in the interval of {-1, 1}. QBR records data information for every path, and uses a simple moving average period to calculate the average path quality. For a period of M, average path quality of every path will be calculated using the most recent M flow data.

For example, let $S = \{0, 0.6, -1, -0.1, 1\}$ represent the data of the last M =5 flows, the average path quality of M flows is (0+0.6-1-0.1+1)/5=0.1.

Now, if a new flow is arrives with accepted quality, then the oldest element will be deleted from S and the set S will be updated to $S = \{0.6, -1, -0.1, 1, 1\}$ and the new path quality would be (0.6-1-0.1+1+1)/5=0.3.

Although the first localized QoS routing algorithm was Proportional Sticky Routing (PSR) [8], it has subsequently been shown that CBR and QBR algorithms outperform PSR algorithm [9] [20] in all situations.

## 3. A New Clustering Metric

Previous studies in the context of global QoS routing algorithms have tended to use average path length and the average node degree [3] [13] [21] to measure the level of connectivity. Clustering structures have a significant impact on the performance of various routing protocols [22]. Further topology parameters can help the analysis of network traffic, congestion and critical network issues [23]. However, localized QoS routing algorithms need a suitable topology that balances the distance between any pair of nodes in the topology with the flexibility of a large number of possible shortest candidate paths, which would reduce the blocking probability. For this reason it is desirable to design a new topology metric that accurately captures the balance level of connectivity. We developed a balance clustering metric (BCM) which is a practical clustering accuracy metric. The basic idea of BCM involves calculating the distances between any pair of nodes and then computing a s*tandard deviation path length* which subjectively specifies how tightly the nodes are clustered throughout the topology. To the best of our knowledge the proposed metric has not been analyzed before, whether in the context of routing algorithms or topology based models.

Let $V$ be a set of topology nodes and L be the set of links in a topology $G(V,L)$, we calculate the BCM metric using path length for each pair of nodes $\Lambda_i$, let $\widehat{M}$ be the path matrix which stores the number of hops along the direct path between source and destination pairs, $\widehat{M}$ is the shortest path length among all node pairs and the diagonal elements of $\widehat{M}$ are zero. The diagonal elements are all zero $M_{ii} = 0$, and off-diagonals contain the number of hops (distance) along the path connecting each pair of nodes $i \neq j$ is $v_i$.

Therefore the BCM metric is calculated using the *standard deviation of the row distances* as follows:

$$BCM = \sqrt{\frac{\sum_{i=1}^{N} \left(\Lambda_i - \overline{m}\right)^2}{N-1}} \qquad (4)$$

Where, $\Lambda_i$ is the sum of rows or columns in the path matrix and $\overline{m}$ is the arithmetic mean. The sum of row distances for each node is calculated by:

$$\Lambda_i = \sum_{i=1}^{N} v_i \qquad (5)$$

The arithmetic mean is $\overline{m} = \sum_{i=1}^{N} \Lambda_i / N$, the BCM metric of the whole topology is:

$$BCM = \sqrt{\frac{\sum_{i=1}^{N} \left( \sum_{i=1}^{N} v_i - \left( \sum_{i=1}^{N} \Lambda_i \Big/ N \right) \right)^2}{N-1}} \qquad (6)$$

where N is the number of nodes. The BCM metric can well reflect the significance of the structure in the topology. The smaller the BCM the more significant the structure, if the BCM=0 this means the distance between each source-destination pair is the same; this is considered to be a balanced topology.

We can calculate the average path length ($\overline{h}$) by summing the non zero elements of the path matrix and dividing by the number of non zero elements.

$\overline{h} = T / N(N-1)$ where T is the sum of off-diagonal elements and N is the number of nodes.

We can obtain the degree and clustering coefficients topology characteristics using its adjacency matrix $\widehat{A}$. Any network topology $G$ with $N$ nodes can be represented by its adjacency matrix $\widehat{A}$ with $N \times N$ elements $A_{ij}$, whose value is $A_{ij} = Aji = 1$ if nodes $i$ and $j$ are connected and 0 otherwise.

For example, we consider a simple topology depicted in Figure 1, in which the number of nodes N=6 and the number of links L=9.
The diagonal elements are all zero and off diagonals contain the number of hops along the path connecting each pair of nodes. In fact the diameter of this topology is 2 hops; therefore the largest value for any element of L can be 2.

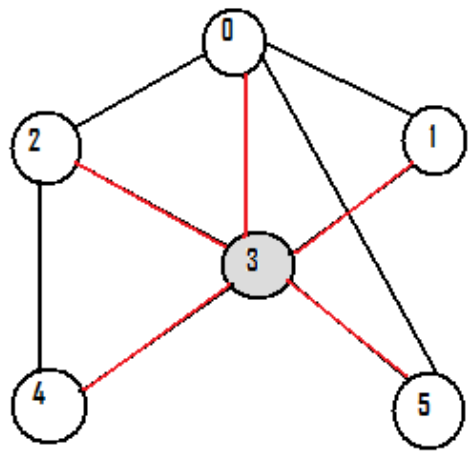

**Figure 1 Illustration of BCM Metric**

The average path length ($\bar{h}$) is 1.4 the clustering coefficient is 0.9133 and the balance clustering metric (BCM) is therefore is 1.265

| $v_0$ | $v_1$ | $v_2$ | $v_3$ | $v_4$ | $v_5$ | | $\Lambda_i$ | | BCM |
|---|---|---|---|---|---|---|---|---|---|
| 0 | 1 | 1 | 1 | 2 | 1 | | 6 | | |
| 1 | 0 | 2 | 1 | 1 | 2 | = | 7 | = | 1.265 |
| 1 | 2 | 0 | 1 | 2 | 2 | | 8 | | |
| 1 | 1 | 1 | 0 | 1 | 1 | | 5 | | |
| 2 | 1 | 2 | 1 | 0 | 2 | | 8 | | |
| 1 | 2 | 2 | 1 | 2 | 0 | | 8 | | |

## 4. Localized BCM Network Model

We developed a simple model using a BCM metric that demonstrates that can be rewired to increase the connections that the network topology accepts. Thus, we assume that the topology is static in the sense that although links can be rearranged, the number of nodes is fixed throughout the forming process. However, our goal is not to provide a thorough evaluation of topology model design, rather we aim to provide an accurate topology structure that maintains best performance for localized routing algorithms.
It is necessary to develop a simple method for network topology in which the level of connectivity can be adjusted in a controlled manner. This method incorporates the addition of new links, the removal of some existing links and rewiring without altering the number of nodes or links in the topology. We thus examine the path matrix form, in which the elements of path matrix $\widehat{M}$ are equal to the distance separating nodes or 0 if no path exists.

Thus, starting from the initial topology structure, it is necessary to monitor changes of topology by removing links for lower distance between existing node pairs and reconnecting to larger node distance pairs in agreement with the level of connectivity: removing the link $\min \Lambda_i$ to $\min v_i$ connected nodes and adding a link to $\max \Lambda_i$ to $\max v_i$ node pairs.

This method is quite natural, since it is possible to make several rewiring changes to decrease the BCM metric and then decrease the average path length significantly. On the other hand, several rewired links will not crucially change the local clustering property of the network. If we consider the example described in the previous section, the number of nodes N=6, the number of links L=9, the clustering coefficient is 0.91 and $BCM = 1.265$, as illustrated in the path matrix:

| $v_0$ | $v_1$ | $v_2$ | $v_3$ | $v_4$ | $v_5$ | | $\Lambda_i$ | | BCM |
|---|---|---|---|---|---|---|---|---|---|
| 0 | 1 | 1 | 1 | 2 | 1 | | 6 | | |
| 1 | 0 | 2 | 1 | 1 | 2 | = | 7 | = | 1.265 |
| 1 | 2 | 0 | 1 | 2 | 2 | | 8 | | |
| 1 | 1 | 1 | 0 | 1 | 1 | | 5 | | |
| 2 | 1 | 2 | 1 | 0 | 2 | | 8 | | |
| 1 | 2 | 2 | 1 | 2 | 0 | | 8 | | |

We remove the link between nodes $v_3 \rightarrow v_0$ which is the lower $\Lambda_i$ and add a link between nodes $v_5 \rightarrow v_4$ which is the larger $\Lambda_i$.

We repeat the rewiring process according to the level of connectivity, the following example demonstrates three rewiring changes and the BCM value has decreased from 1.265 to 0. The path matrix and topology representation is then modified as follows:

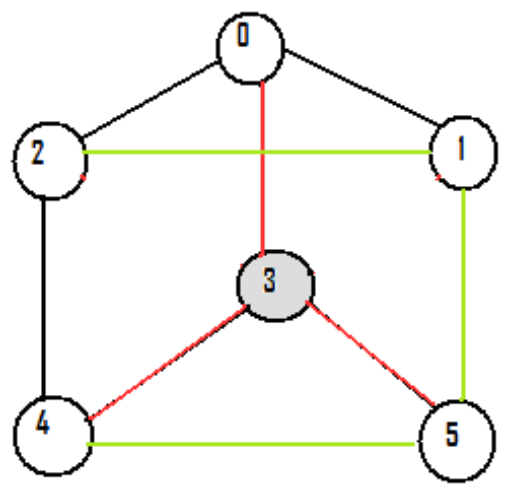

**Figure 2 BCM network model**

| $v_0$ | $v_1$ | $v_2$ | $v_3$ | $v_4$ | $v_5$ | | $\Lambda_i$ | | BCM |
|---|---|---|---|---|---|---|---|---|---|
| 0 | 1 | 1 | 1 | 2 | 2 | | 7 | | |
| 1 | 0 | 1 | 2 | 1 | 2 | = | 7 | = | 0 |
| 1 | 1 | 0 | 2 | 2 | 1 | | 7 | | |
| 1 | 2 | 2 | 0 | 1 | 1 | | 7 | | |
| 2 | 1 | 2 | 1 | 0 | 1 | | 7 | | |
| 2 | 2 | 1 | 1 | 1 | 0 | | 7 | | |

We noted that the average path length is 1.4 and the BCM metric decreased to 0.

# 5. Performance Evaluation

We evaluate the performance of the proposed method using the localized QBR scheme and the localized CBR scheme. In the following, we first describe the simulation model and then compare the performance of the QBR and CBR schemes using the network model described in the previous Sections.

## 5.1. Simulation Model

We conducted extensive simulations implemented under OMNeT++ [24] [25] in order to test the performance of the localized QoS routing algorithms proposed in the previous Sections, according to the level of network connectivity. Consequently, our simulation experiments consider a range of topologies with similarities and differences in their important parameters, such as average path length, node degree and diameter, including standard deviation path length. We comment on similarities and differences between the trends in each topology.

We consider random topologies with relatively different levels of connectivity to evaluate the effects of having multiple shortest candidate paths between pairs of nodes. In addition, we consider ISP topology which has been classified as a sparsely connected topology. The random topologies were generated using the BRITE generator [26] using Waxman's model [27]. We also consider the scale-free topology which was generated on top of the BRITE generator [26] using the Barabási-Elbert model [16].

## 5.2. Traffic Generation

All links are assumed to be symmetrical, bidirectional and have the same capacity C (C=150 Mbps) in each direction. We further assumed that the network topology remains fixed throughout each experiment in the simulation; hence we do not model the effects of link failures.

Flows arrive to each source node according to a Poisson distribution with rate $\lambda$=1 and destination nodes are selected randomly by uniform distribution. Flow duration is exponentially distributed with mean value $1/\mu$, while flow bandwidth (QoS requested) is uniformly distributed within [0.1 – 2MB] interval. Following [11] , the offered network load is $\rho = \lambda Nhb/\mu LC$, where N is the number of nodes, L is the number of links in the network, b is the average bandwidth required by a flow, and h is the average path length (averaged across all source-destination pairs). The parameters used in the simulation for CBR are MAX_CREDITS=5 and Φ=1. Blocking probabilities are calculated based on the most recent 20 flows.

A set of candidate paths is chosen such that for each source-destination pair in a selected network topology that is chosen, the candidate set consists of paths that have at most one hop more than the minimum number of hops. All results were collected for all simulation experiments from at least 2,000,000 connection request arrivals, and the results were collected after 200,000 connection requests to allow a steady state to be reached. Each simulation experiment was repeated 20 times with different seeds; the 95% confidence intervals were computed and found to be extremely tight, such that in most figures only mean values of the results are presented.

### 5.3. Performance Metrics

The performance metrics used to measure the performance of the algorithms are flow blocking probability and bandwidth blocking probability. Flows are rejected when one of the links along the path from source to destination does not satisfy the requested bandwidth. The blocking probability is defined as:

$$\text{Flow blocking probability}=\frac{\text{Number of rejected requests}}{\text{Number of requests arriving}} \qquad (7)$$

We use also the notion of bandwidth blocking probability to solve discrimination against flows with large bandwidth requirements.

The bandwidth blocking probability is defined as:

$$\text{Bandwidth blocking probability} = \frac{\sum_{i\in B} bandwidth(i)}{\sum_{i\in c} bandwidth(i)} \qquad (8)$$

Here, B is the set of blocked paths and C is the set of total requested paths, and bandwidth (i) is the requested bandwidth for path i.

## 6. Simulation Results

### 6.1. Performance Prediction of Localized Routing Algorithms

The localized QoS routing algorithms depend on the underlying network topology. To study the effects of topology on the performance, we evaluate four random topologies with similar size n = 18 and number of links L=58 under the same traffic load using the BCM metric described in Section 3.

The main parameters and characteristics of the configurations are listed in Table 1.

Table 1 Characteristics of random topologies

| Name | NODES | LINKS | Avg. degree | Avg. path length | BCM |
|---|---|---|---|---|---|
| Network1 | 18 | 58 | 3.22 | 2.32 | 2.09 |
| *Network2* | 18 | 58 | 3.22 | 2.346 | 2.27 |
| *Network3* | 18 | 58 | 3.22 | 2.43 | 5.77 |
| *Network4* | 18 | 58 | 3.22 | 2.5 | 6..92 |

We have shown that the balance clustering metric BCM is increasing based on the level of connectivity. A lower BCM typically implies a dense topology with balanced path lengths and more flexibility in selecting routes, whereas a larger BCM implies lower connectivity. The differences in the BCM metric shown in Figure 3 reveal a significant influence regarding how well localized algorithms perform.

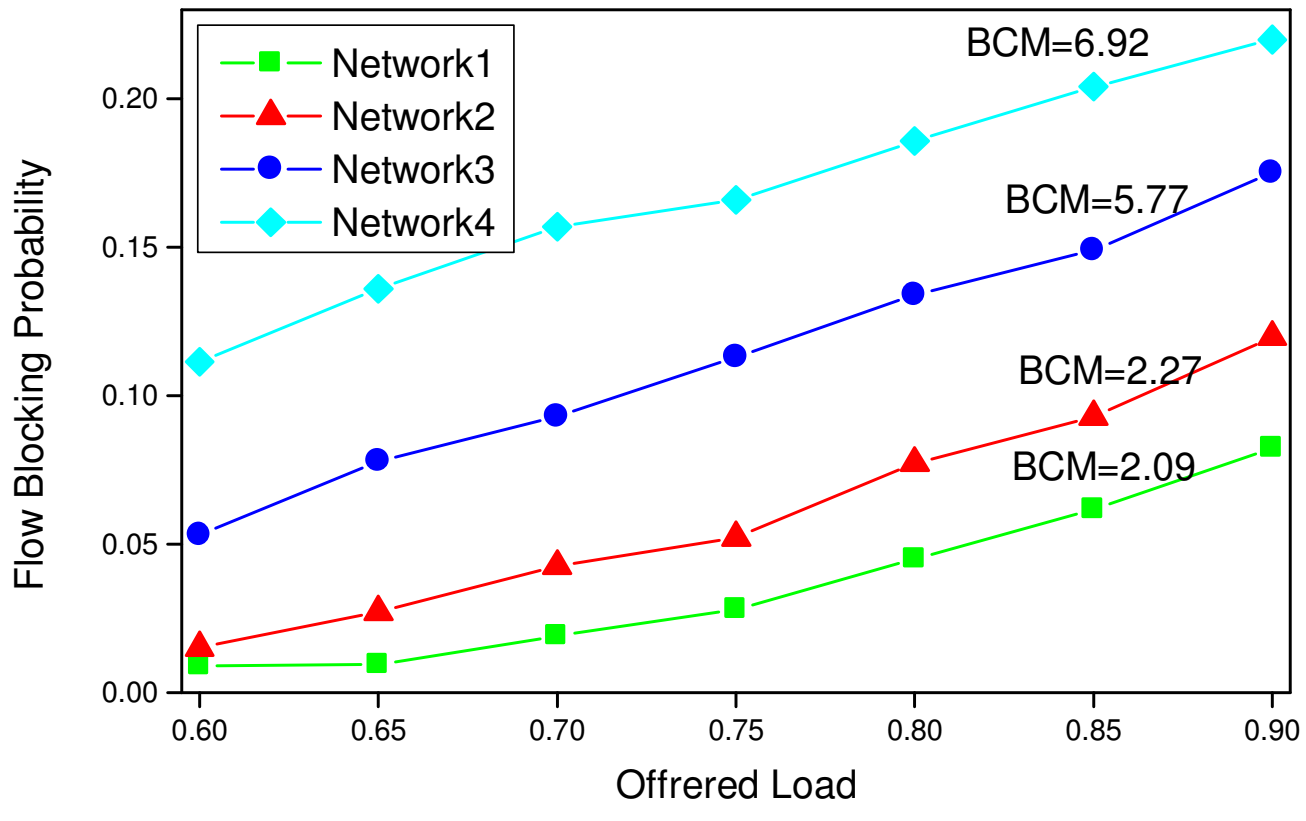

(a) CBR (Flow Blocking Probability)

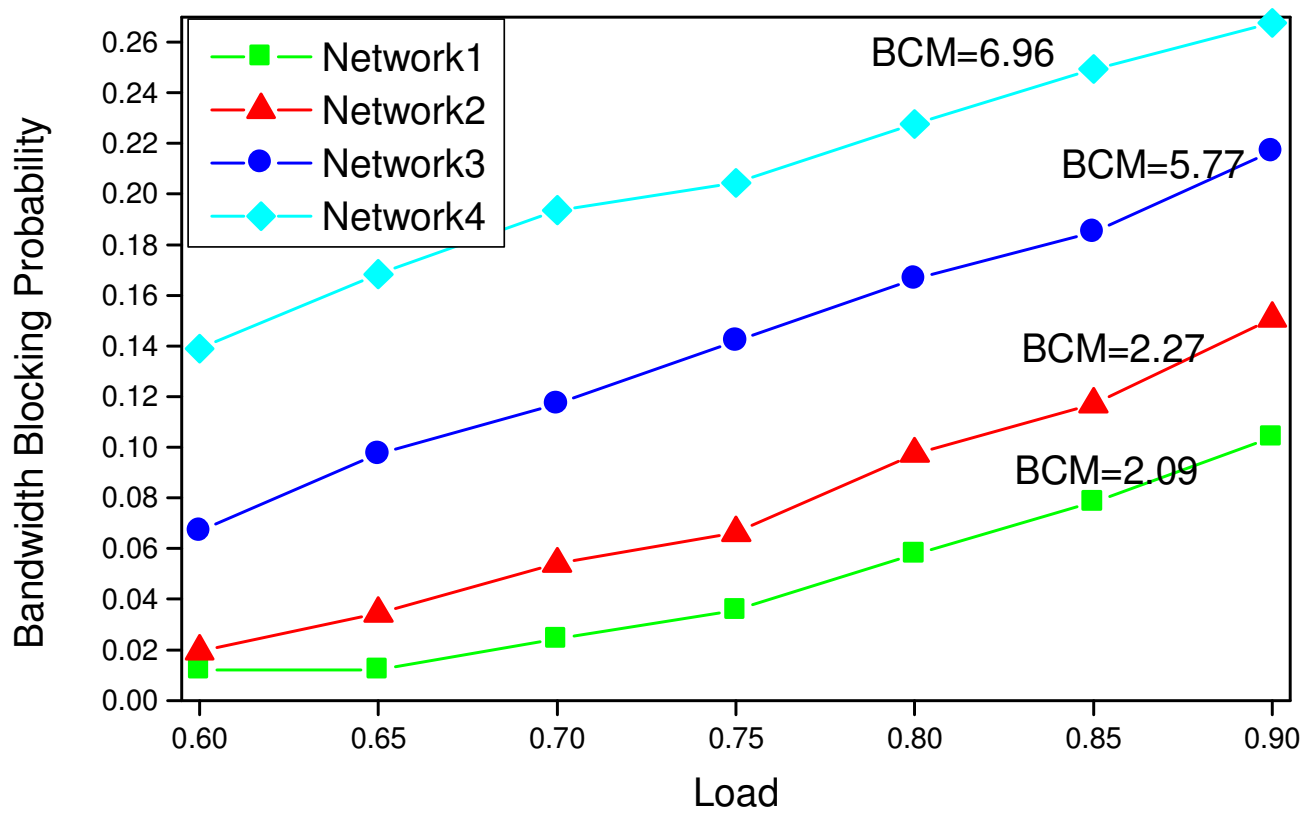

Figure 3 Comparison of random topologies of CBR scheme using flow and bandwidth blocking probabilities

Figure 3 plots the flow blocking probability and bandwidth blocking probability as a function of offered load, using the CBR algorithm we compare between topologies using the same load. We increase the offered load by changing the mean holding time. Figure 3 shows the performance using both metrics. The performance of localized CBR strongly depends on the network topology measured by the value of the BCM metric; the random topology with higher connectivity has the flexibility of a large number of possible routes which reduces the blocking probability.

However, the performance for the random topology with a lower connectivity degrades, which is more likely when most pairs of nodes have a larger path length of the network.

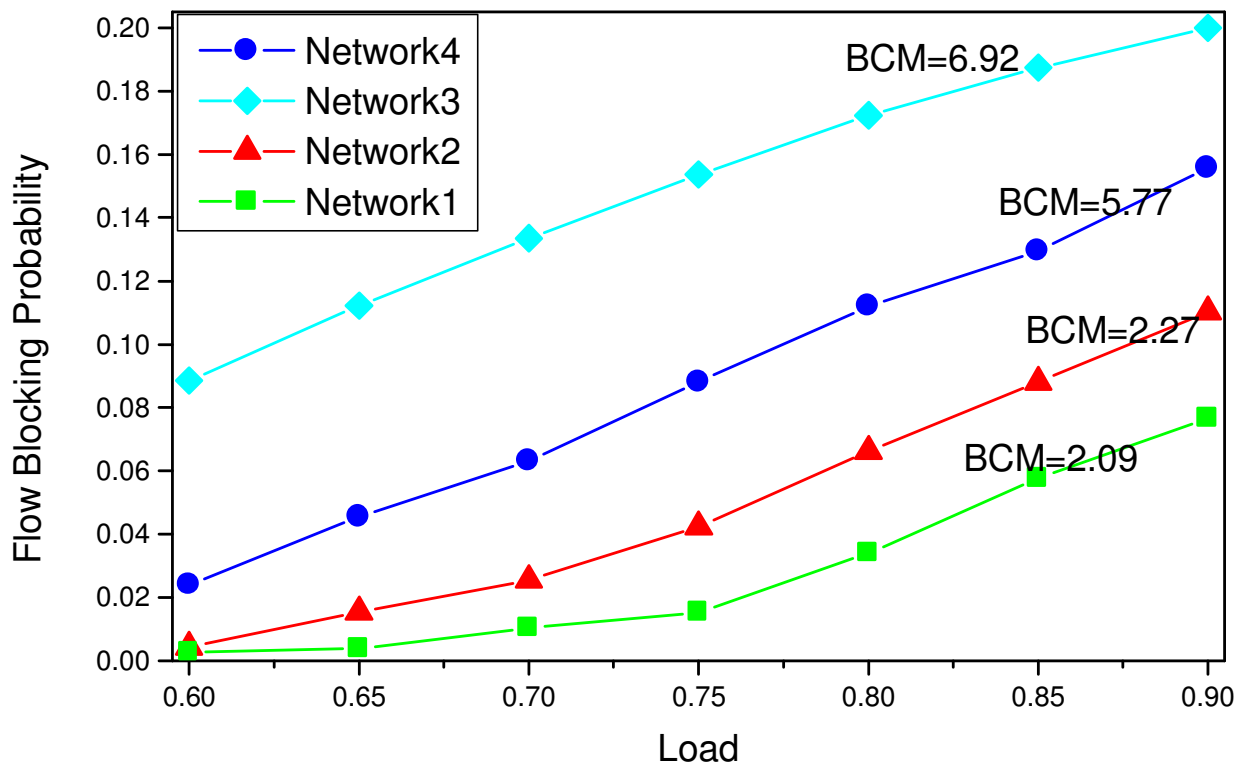

(a) QBR (Flow Blocking Probability)

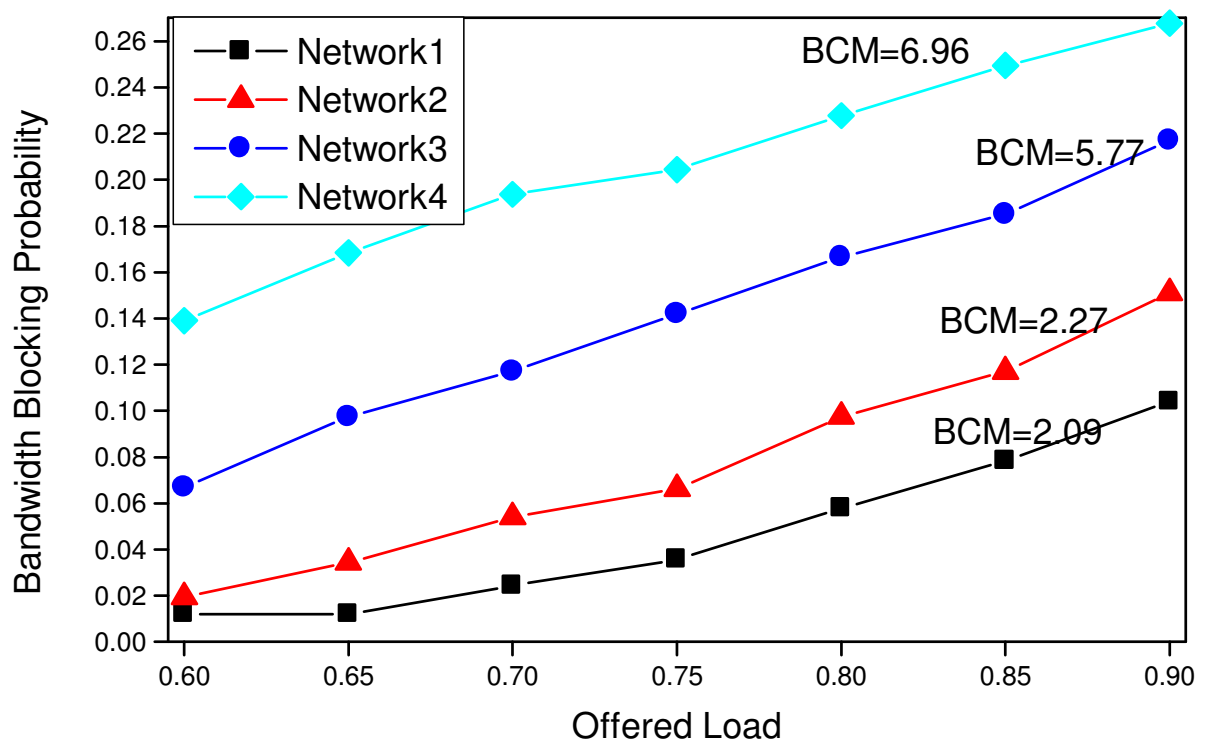

(a) QBR (Bandwidth Blocking Probability)

Figure 4 Comparison of random topologies of QBR scheme using flow and bandwidth blocking probabilities

Similarly, Figure 4 compares the same topologies using both performance metrics for the QBR algorithm, the performance of the QBR algorithm depends on the level of connectivity.
We also evaluated the performance of sparse ISP topology, as depicted in Table 2. However, the characteristic of ISP topology has relatively lower connectivity compared to random topologies. Table 2 lists the main parameters of the proposed topologies.

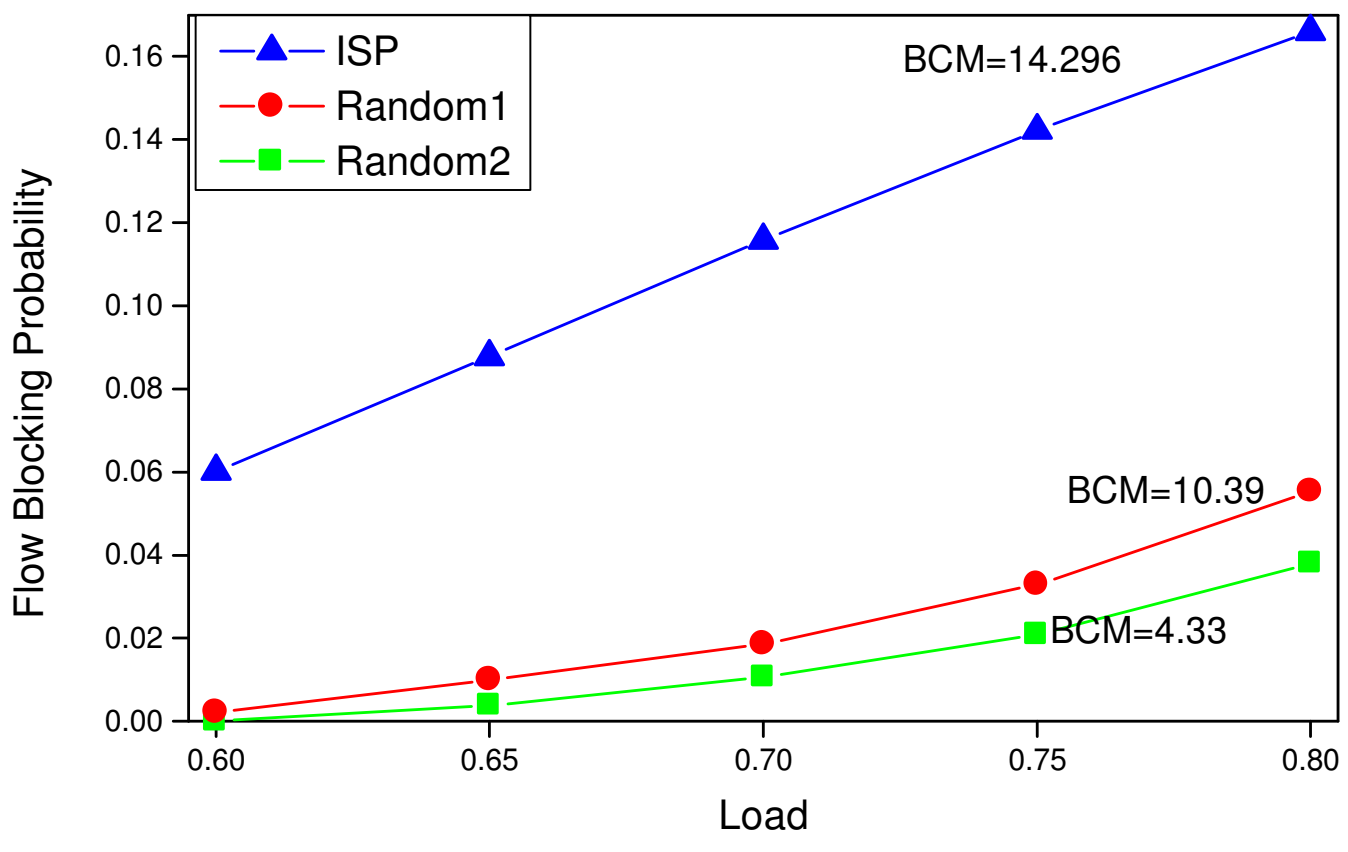

(a) CBR

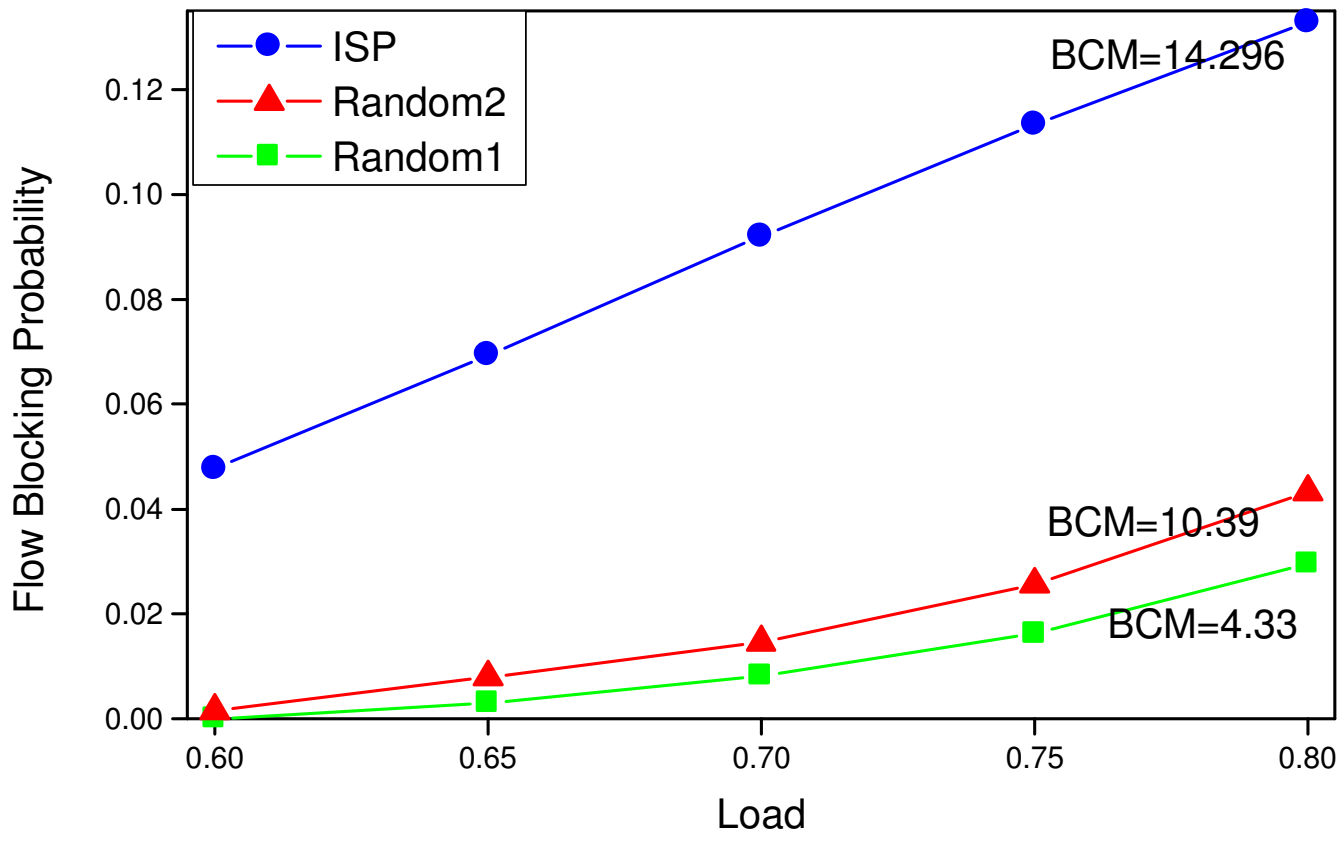

(b) QBR

Figure 5 Comparison of random and ISP topologies using flow blocking probability

Table 2 Characteristics of random and ISP topologies

| Name | Nodes | Links | Avg. degree | Avg. path length | BCM |
|---|---|---|---|---|---|
| ISP | 32 | 108 | 3.375 | 3.177 | 14.296 |
| *Random1* | 32 | 122 | 3.8125 | 2.494 | 4.329 |
| *Random2* | 32 | 122 | *3.8125* | 2.416 | 10.388 |

Figure 5 shows the comparison between ISP and random topologies. It can be seen that the performance degrades as the value of the BCM metric increases. Figure 5 shows that localized routing algorithms usually discriminate against alternative paths in which the algorithms prefer the shortest paths. It has been shown that the localized routing algorithms typically perform well in dense networks (i.e. Random 32 topology) which have the flexibility of a large number of routes between each source and destination.
On the other hand, these algorithms are hard to use scarce resources in sparsely connected ISP topologies as they consume more resources to accommodate future flows.
More generally, due to the growth of the Internet and increasing demand for predictable performance, the simulation results observed an ability to predict topology. Using the BCM metric it is possible to predict, between topologies, which are likely to give better performance.

## 6.2. Network based Routing Model

Figure 6 shows simulation results of the ISP topology which continuously adds links and removes existing links to the topology using the method described in Section 4 until a rich connectivity is achieved.

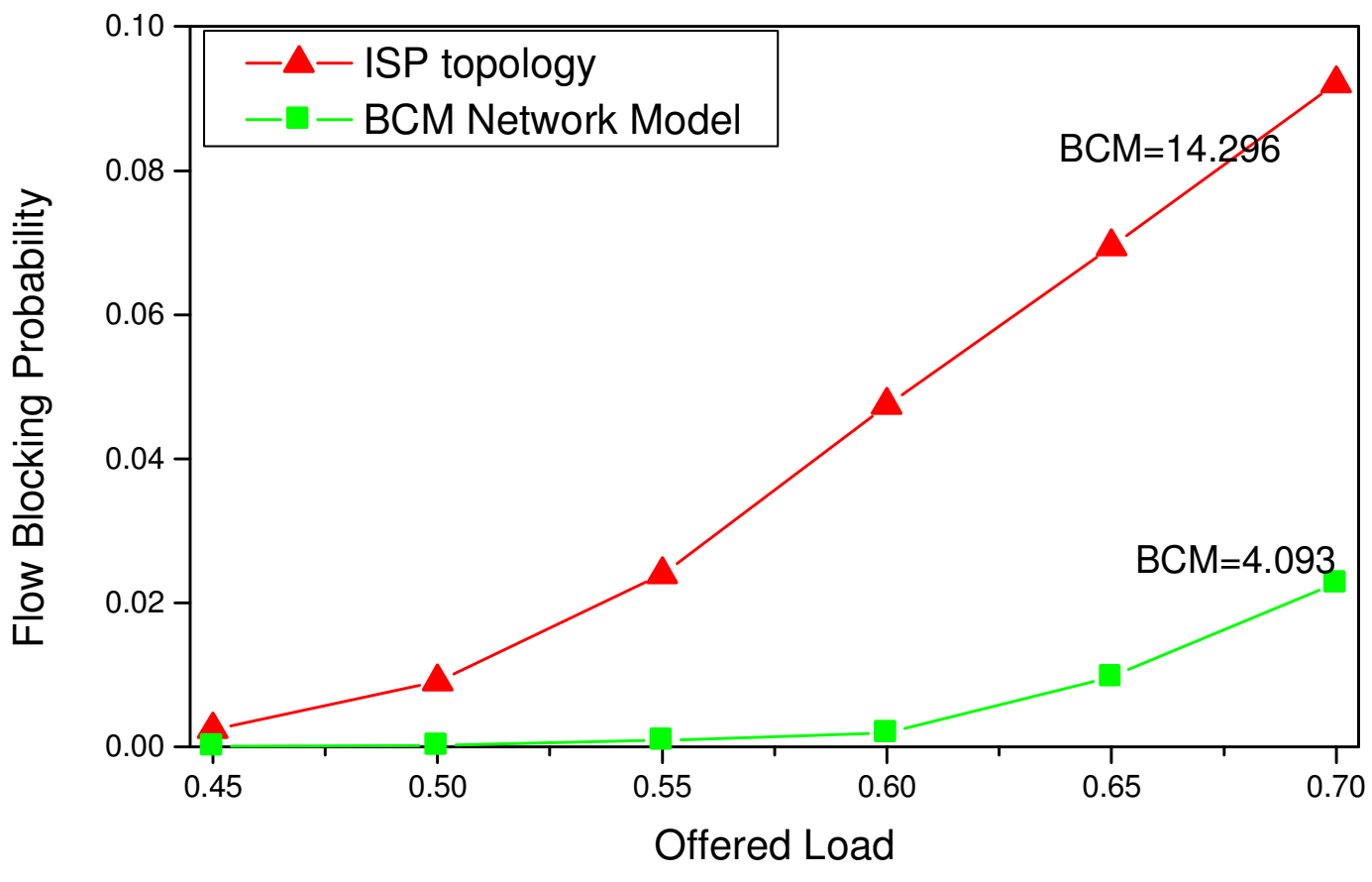

Figure 6 Comparison of the BCM network model with ISP topology

Figure 6 compares the performance of the ISP topology and BCM topology. It can be observed that the BCM topology enhances the performance using the same number of nodes and links of the topology. Using larger paths in sparse topologies tends to consume more resources for future flows causing performance degradation. In this form it is of negligible importance when load is low, as there are still sufficient resources to route future flows.

Figure 7 compares the performance of the QBR algorithm under the scale-free topology and BCM topology, the performance is enhanced by changing the topology to BCM topology.

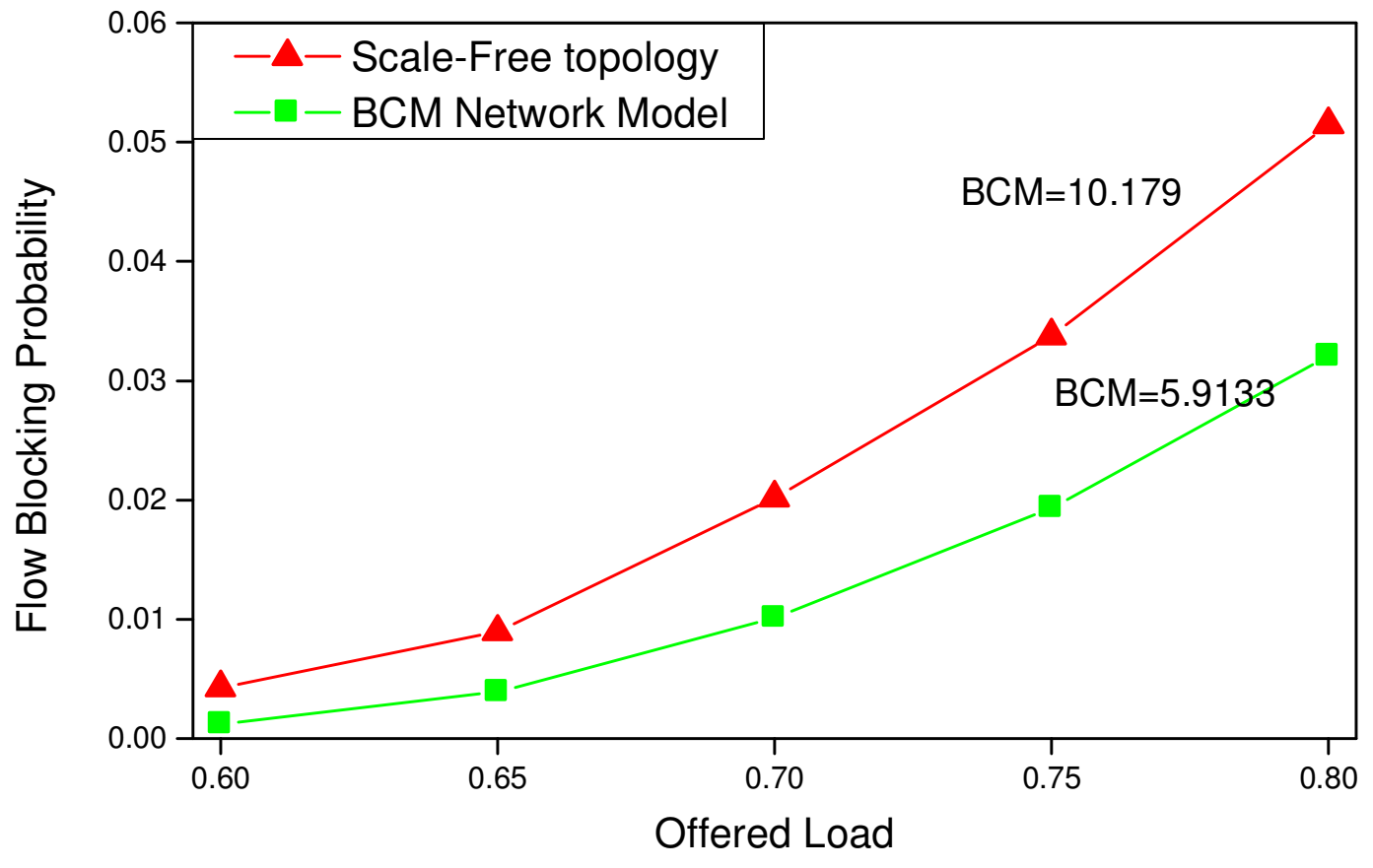

Figure 7 Comparison of the BCM network model with scale-free topology

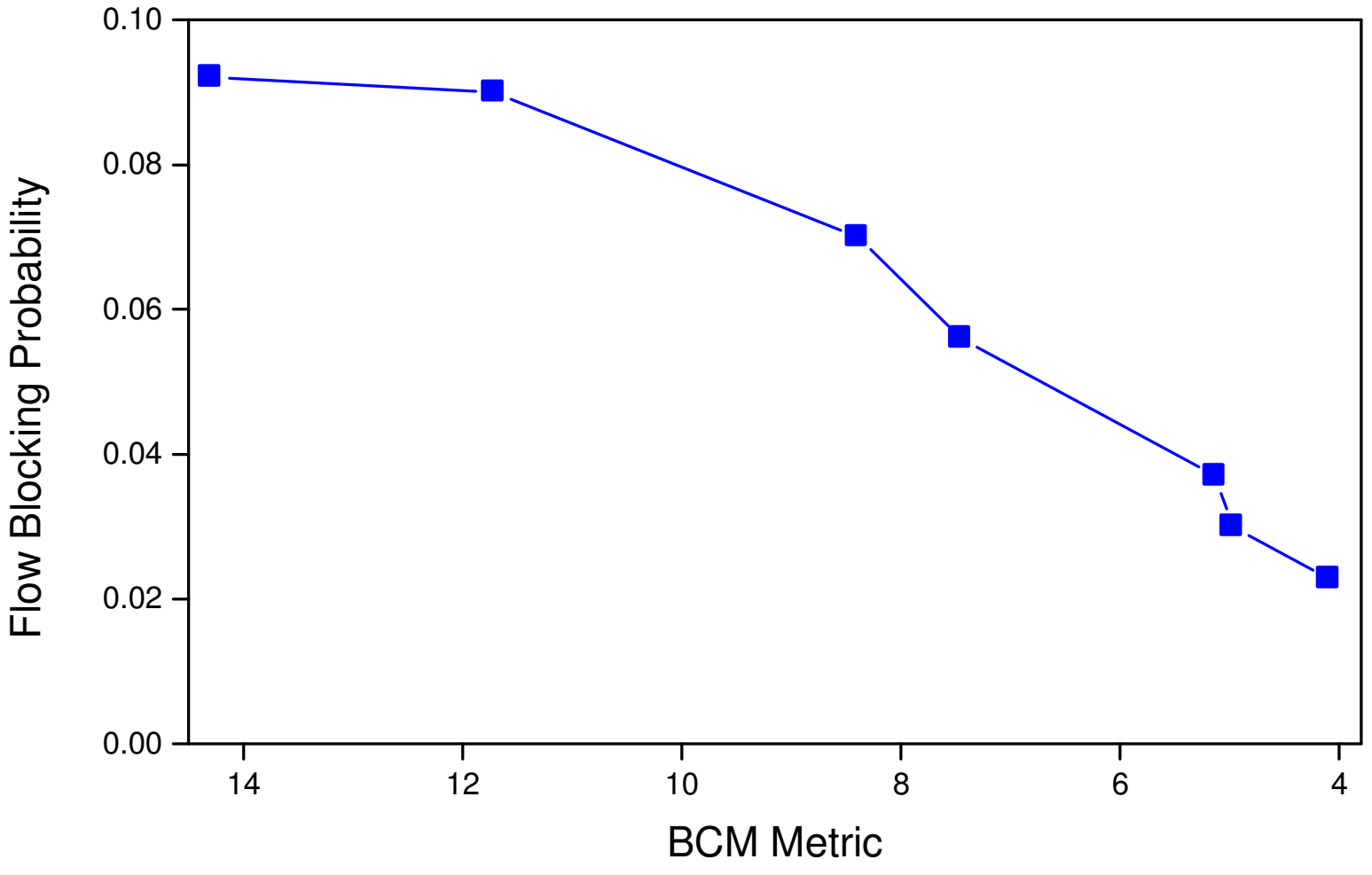

(a) Flow Blocking Probability

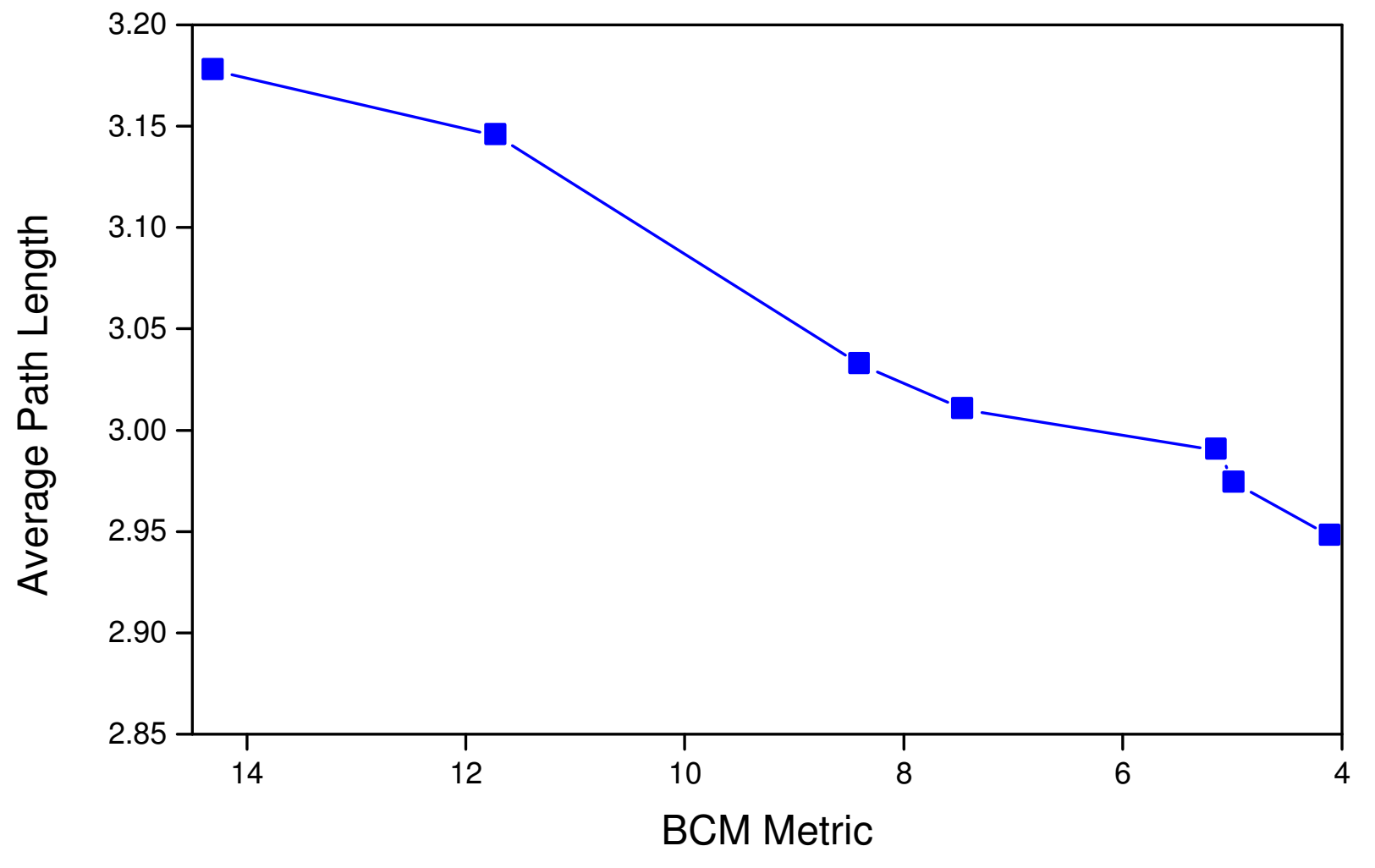

(b) Average Path Length

Figure 8 Flow blocking probability and average path length

Figure 8 (a) observed a decreasing blocking probability as a function of decreasing the BCM metric. Figure 8(b) observed a decrease in the average path length in the BCM network which significantly satisfies the small world property and maintaining the same size of the original network.

## 6.3. Varying Non-Uniform Traffic

The destination nodes in the simulation experiments have been selected uniformly.

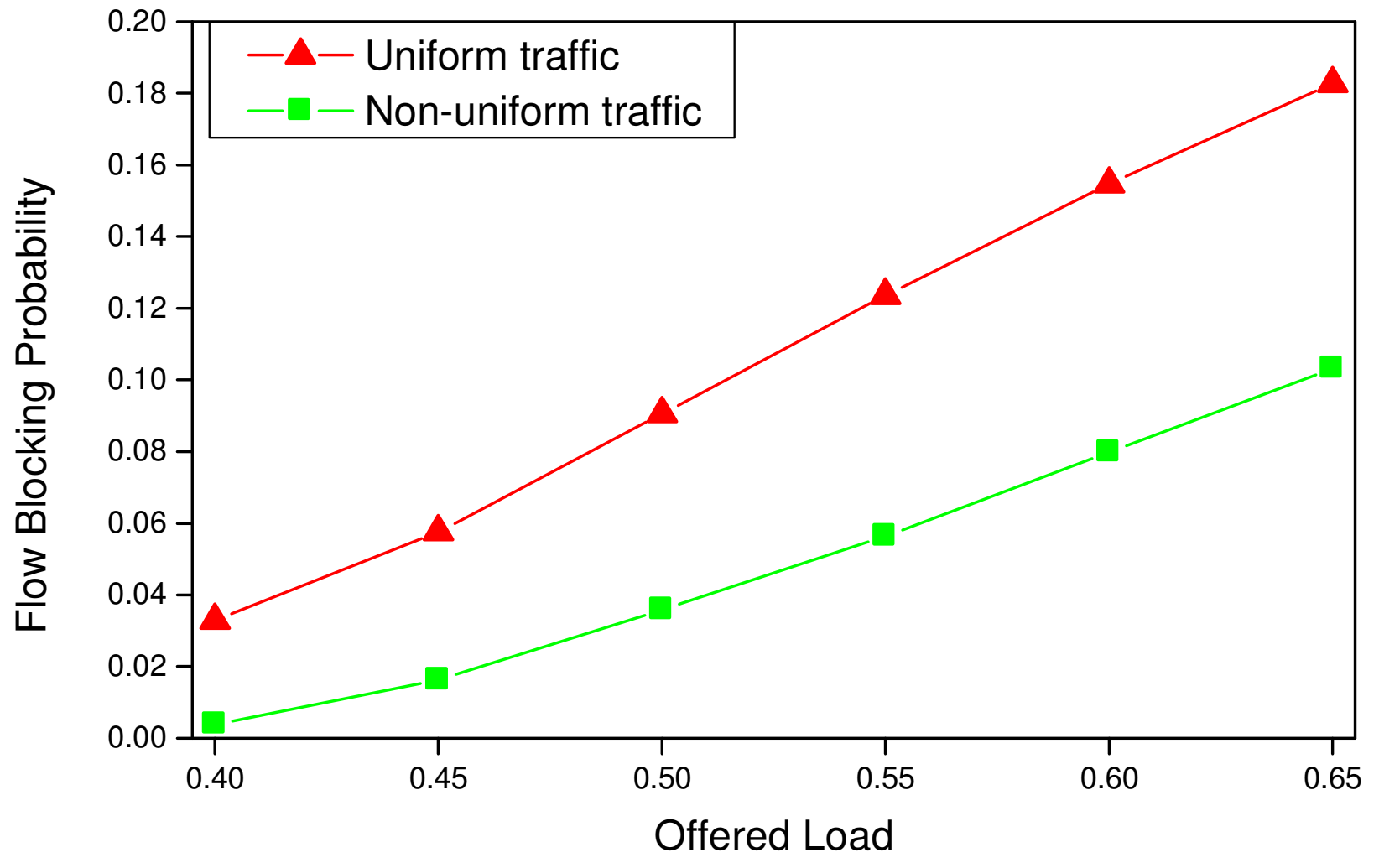

Figure 9 Impact of varying non-uniform traffic

However, in realistic networks the source nodes may receive more traffic demands, especially in the case of the communications in the sub-networks, which usually receive more demands than the communications across sub-networks.
The authors in [28] emphasized that the global uniform end-to-end IP QoS solution is not realistic. For these reasons, the scale-free topology has been modified and has been virtually divided into two sub-networks. The BCM metric of the sub-network1 is 4.22, the BCM metric of the sub-network2 is 4.5 and the BCM for the network topology is 8.72.
We use a varying of non uniform traffic across the scale-free topology; the traffic demands routed inside the sub-network are three times higher than traffic routed between sub-networks.
In Figure 9 flow blocking probability is plotted against different load conditions using uniform and non-uniform traffic for scale-free topology. It can be noticed that the localized routing algorithms perform extremely well under non-uniform traffic compared with uniform traffic. Hence, by formulating the correlated traffic in the building blocks of realistic topologies using the BCM metric, we are effectively reducing the blocking probability to obtain superior performance.

## 7. Conclusions

This paper investigated a new clustering metric based on localized QoS routing algorithms taking path length into consideration. The BCM metric showed to be a good indicator and can be used to predict which network gives better performance. We developed a BCM network model using a BCM metric that rewired to introduce increasing the performance. The results of the model have small characteristic path length under random and complex network topologies. The simulation results obtained show that the localized QoS routing algorithms typically perform better for highly connected networks where they are likely to be able to balance the load over the set of minimum hop candidate paths.
However, the simulation results show that the localized QoS routing algorithms being hard to route scarce resources over sparse topologies. We generally suggest that the localized QoS routing algorithms should distribute the load according to connectivity levels to avoid congestion and be able to model the diversity seen in current realistic networks, such as the Internet.

## REFERENCES

[1] B. Peng, A. Kemp, and S. Boussakta, "Impact of network conditions on QoS routing algorithms," in *3rd IEEE Consumer Communications and Networking*, pp. 25-29, 2006.

[2] N. Ansari, G. Cheng, and N. Wang, "Routing-oriented update scheme (ROSE) for link state updating," *IEEE Transactions on Communications,* vol. 56, pp. 948-956, 2008.

[3] A. Shaikh, J. Rexford, and K. Shin, "Evaluating the Impact of Stale Link State on Quality-of-Service Routing," *IEEE/ACM Transactions on Networking,* vol. 9, pp. 162-176, 2001.

[4] B. Fu, F. A. Kuipers, and P. V. Mieghem, "To update network state or not?" in *The 4th international telecommunication networking workshop on QoS in multiservice IP networks*, Feb 2008.

[5] E. W. Zegura, K. L. Calvert, and S. Bhattacharjee, "How to model an inter-network," in *Proceedings of IEEE INFOCOM*, pp 494-602, March 1996.

[6] V. Paxson and S. Floyd, "Why we don't know how to simulate the Internet," in *Proceedings of the 1997 Winter Simulation Conference*, 1997.

[7] R. Pastor-Satorras and A. Vespignani. Immunization of complex networks. *Phys. Rev. E*, 65, 2002.

[8] S. Nelakuditi, Z. L. Zhang, R. Tsang and D. Du, Adaptive Proportional Routing: a Localized QoS Routing Approach, IEEE/ACM Transactions on Networking 10 (2002), pp. 790-804.

[9] S. Alabbad, M. E. Woodward. "Localized Route Selection for Traffic with Bandwidth Guarantees", Simulation, Transactions of the SCS, Vol.83, No.3, pp. 259-272, March 2007.

[10] A. H. Mohammad. "New Localized Call Admission Control Algorithms in Communication Networks with Quality of Service Constraints", International Journal of Computer Science and Network Security, Vol.10, No.11, pp. 125-131, November 2010.

[11] D. Awduche, J. Malcolm, J. Agogbua, M. O'Dell and J. McManus, "Requirements for Traffic Engineering over MPLS," *RFC-2702,* September 1999.

[12] X. Masip-Bruin, E. Marin-Tordera, M. Yannuzzi, R. Serral-Garcia and S. Sanchez-Lopez, "Reducing the effects of routing inaccuracy by means of prediction and an innovative link-state cost" *IEEE Communications letters*, vol. 14, no. 5, May 2010.

[13] K. Kowalik and M. Collier, "Should QoS routing algorithms prefer shortest paths?", *IEEE International Conference on Communications*, Anchorage Alaska, May 2003.

[14] R. Gawlick, A. Kamath, S. Plotkin, and K. Ramakrishnan, "Routing and Admission Control in General Topology Networks," *Technical Report STAN-CS-TR-95-1548,* 1995.

[15] S. H. Strogatz, "Exploiting complex networks," *Nature*, vol. 410, pp 268-276, March 2001.

[16] R. Albert and A-L. Barabasi, "Statistical mechanics of complex networks," *Review of Modern Physics*, vol. 74, pp.47-91, January 2002.

[17] Z. Wang, A. Scaglione, R. Thomas, "Generating Statistically Correct Random Topologies for Testing Smart Grid Communication and Control Networks". *IEEE Transactions on Smart Grid*, vol. 1, no. 1, June 2010.

[18] P. Giabbanelli, "Impact of complex network properties on routing in backbone networks". In *Proceedings of the IEEE Globecom 2010 Workshop on Complex and Communication Networks*

[19] L. Li, D. Alderson, R. Tanaka, J. C. Doyle, and W. Willinger. Towards a theory of scale-free graphs: Definition, properties and implications. *Internet Mathematics*, 2:431–523, 2005.

[20] A. H. Mohammad, M. E. Woodward. "Localized Quality Based QoS Routing", Proc. Of International Symposium on Performance of Computer and Communications systems IEEE press, SPECTS June 2008

[21] K. Kowalik, M. Collier, "Connectivity Aware Routing - a Method for Finding Bandwidth Constrained Paths Over a Variety of Network Topologies". *IEEE International Symposium on Computers and Communications*, pp. 392-398 July 2003.

[22] P. Fraigniaud, "A New Perspective on the Small-World Phenomenon: Greedy Routing in Tree-decomposed Graphs," in *Proceedings of 13th Annual European Symposium on Algorithms (ESA)*, pp. 791-802, 2005.

[23] P. Mahadevan, D. Krioukov, K. Fall and A. Vahdat. "Systematic Topology Analysis and Generation using Degree Correlations," in *Proceedings of ACM SIGCOMM*, pp. 135-146, 2006.

[24] A. Varga, "OMNeT++ Community Site," in *URL:http://www.omnetpp.org*, 2010.

[25] A. Varga, "The OMNeT++ Discrete Event Simulation System," in *the European Simulation Multi-conference*, Prague, Czech Republic, 2001.

[26] A. Medina, A. Lakhina, I. Matta, and J. Byers, "Brite: An Approach to Universal Topology Generation," *The International Workshop on Modeling Analysis and Simulation of Computer and Telecommunications Systems*, Cincinnati, Ohio, 2001.

[27] B. M. Waxman, "Routing of multipoint connections," *IEEE J. Select. Areas Communications,* vol. 6(9), pp. 1617-1622, 1988.

[28] J. Soldatos, E. Vayias, and G. Kormentzas, "On the building blocks of quality of service in heterogeneous IP networks," *IEEE Communications Surveys and Tutorials,* vol. 7, 2005.

**Author**

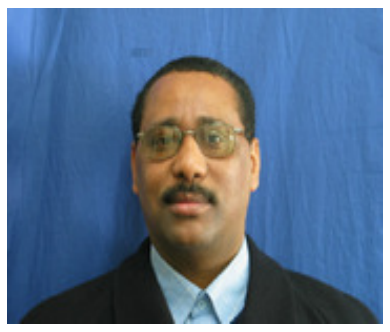

Abdulbaset Mohammad is a Ph.D candidate and a member of networks and performance engineering research group at the University of Bradford. He received the B.Sc degree in Communication Engineering from Libya in 1990 and M.Sc degree from Technical University of Budapest in Telecommunication Engineering in 2000 with excellent honours thesis in the field of routing algorithms in ATM networks. He is working as a lecturer in Libya's higher education. His current research interest includes QoS routing algorithms, call admission control algorithms, complex networks and network based routing algorithms. He is a member of the IEEE, a member of the IEEE Communication Society, a member of the SCS and has a recipient of the Libya's higher education scholarship.